\documentclass[conference,a4paper]{IEEEtran}

\usepackage{enumerate}
\usepackage{epstopdf}
\usepackage{cite}
\usepackage{amscd}
\usepackage{dsfont}
\usepackage{latexsym}
\usepackage{array}

\usepackage{tikz}\usetikzlibrary{calc}
\usepackage{tabularx}
\usepackage{lipsum}
\usepackage{epsfig}
\usepackage{placeins}
\usepackage{afterpage}
\usepackage{graphicx,subfigure}
\usepackage{placeins}
\usepackage{amsmath,mathrsfs}
\usepackage{amsthm}
\usepackage{amssymb}  
\usepackage{amsbsy}
\usepackage{color}
\usepackage{float}
\usepackage{algorithm}
\usepackage{booktabs}
\usepackage{algpseudocode}
\usepackage{dblfloatfix}
\usepackage{tabularx}
\usepackage{bm}
\usepackage{pifont}
\usepackage{accents}
\usepackage{multirow}

\theoremstyle{definition}
\newtheorem{rem}{Remark}

\newtheorem{theorem}{Theorem}
\newtheorem{lemma}{Lemma}
\newtheorem{corollary}{Corollary}
\begin{document}
\title{Pixel-Sensitive and Robust Steganography \\ Based on Polar Codes}
\author{
\IEEEauthorblockN{Yujun Ji\IEEEauthorrefmark{1}, Jinsheng Li\IEEEauthorrefmark{1}, Ling Liu\IEEEauthorrefmark{2}, Qi Cao\IEEEauthorrefmark{2} and Tao Dai\IEEEauthorrefmark{1}}
\IEEEauthorblockA{
\IEEEauthorrefmark{1}
College of Computer Science and Software Engineering, Shenzhen University, China}
\IEEEauthorblockA{
\IEEEauthorrefmark{2}
Guangzhou Institute of Technology, Xidian University, Guangzhou, China}
Email: \{jiyujun2023, lijinsheng2020\}@email.szu.edu.cn, \{liuling, caoqi\}@xidian.edu.cn, daitao@szu.edu.cn 
}
\maketitle

\begin{abstract}
    Steganography is an information hiding technique for covert communication. The core issue in steganography design is the rate-distortion coding problem. Polar codes, which have been proven to achieve the rate-distortion bound for any binary symmetric source, are utilized to design a steganographic scheme that can reach the embedding capacity for the Distortion-Limited Sender problem in certain cases. In adaptive steganography, for attack scenarios where each noise element can have different intensities, existing steganographic coding methods fail to resist such attacks. In this paper, we propose a pixel-sensitive and robust steganographic scheme based on polar codes. Our steganographic scheme not only matches the adaptive distortion well but is also robust against sophisticated noise attacks. Futher, it is proven that our scheme achieves the embedding capacity in certain cases. Experimentally, a steganographic scheme can be designed and implemented with a secret message error rate at the $10^{-5}$ level when the attack noise is known to both the sender and the receiver. This demonstrates its significant robustness.

\end{abstract}
\IEEEpeerreviewmaketitle

\section{Introduction}
In recent years, the rapid development of computer networks, including wireless networks, has increased the security risks of information dissemination, drawing more attention to information security \cite{bloch2011physical}. As an information hiding technique for covert communication, steganography secures data transmission in digital networks.  Steganography hides secret messages in ordinary covers (e.g. image, audio and text.), making it difficult for eavesdroppers to notice the presence of secret messages \cite{moulin2005data, fridrich2010steganography, mazurczyk2014steganography}.

Steganalysis is a technique antagonistic to steganography, aiming to detect the presence of secret messages in a carrier. To counter steganalysis, steganography needs to minimize the distortion to the cover caused by embedding messages at a given payload rate, which is known as the Payload-Limited Sender (PLS) problem. The Distortion-Limited Sender (DLS) problem is the dual of the PLS problem, where the goal is to maximize the payload rate given a certain level of distortion caused by embedding messages. The core issue in steganography design is the rate-distortion coding problem, which can be solved using steganographic codes \cite{zhang2008coding,zhang2012efficient,diop2014adaptive}. 
In adaptive steganography, a distortion function that represents the modification distortion weight of each cover element is defined. When the modification of each cover element is independent, the distortion function is additive. Currently, distortion functions can be derived using advanced deep learning techniques \cite{huang2023steganography, huang2023automatic, mo2023reload}. For adaptive steganography, Syndrome-Trellis Codes (STC) \cite{filler2011minimizing} are most widely used, as they can asymptotically approach the bounds for a large range of distortion functions.

Polar coding, introduced by Ar{\i}kan \cite{arikan2009channel}, is a channel coding method that can be proven to achieve the symmetric capacity of Binary Discrete Memoryless Channels (B-DMCs). It has also been extended to the field of source coding \cite{arikan2010source}, \cite{korada2009polar}. In lossy source coding, Korada and Urbanke proved that polar codes can achieve the rate-distortion bound for any binary symmetric source (BSS) \cite{korada2010polar}. Therefore, using polar codes as steganographic codes is highly suitable. Diouf \emph{et al.} were the first to use a SC decoder as a steganographic encoder \cite{diouf2018polar}, and experimentally demonstrated better embedding performance compared to STC. 
Subsequent works \cite{li2020designing, fu2022high, fu2024performance} further optimized and expanded the steganography based on polar codes.

However, in practical applications, the stego sequence may suffer from noise attacks due to various factors, such as adversary-originated noise or transmission through a noisy communication channel, leading to secret information loss. Therefore, robust steganography is required.  The steganographic methods in \cite{diouf2018polar,li2020designing,fu2022high,fu2024performance} cannot withstand noise attacks. 
For attack scenarios where each noise element has the same intensity, Li and Liu proposed robust schemes based on polar codes for the constant distortion (where the modification distortion weight of each cover element is the same) \cite{li2021robust}. 
Yao \emph{et al.} proposed reliable robust adaptive steganographic coding based on nested polar codes for the adaptive distortion \cite{yao2024reliable}. 
In this paper, for generalized attack scenarios where each noise element can have different intensities, we propose a pixel-sensitive and robust steganographic scheme based on polar codes. Our steganographic scheme not only matches the adaptive distortion well but is also effective against more sophisticated noise attacks. More suitable construction algorithms and better-performing decoders of polar codes are employed in our scheme. We theoretically prove that our scheme achieves the embedding capacity in certain cases.

All random variables (RVs) are denoted by capital letters, and their realizations are denoted by the corresponding lowercase letters. $\left [ N \right ] $ denotes the set $\left \{ 1,2,\dots,N  \right \} $. We use the notation $x^{j}_{i}(i\le j)$ as a shorthand for the vector $(x_i,..., x_j)$, which is a realization of the RVs $X^{j}_{i}=(X_i,..., X_j)$. 
The vector($x_i:i\in\mathcal{A}$) is denoted by $x_{\mathcal{A}}$. The Shannon entropy of $X$ is denoted by $H(X)$, and the binary entropy function is denoted by $h_2 (\cdot )$. For a set $\mathcal{I}$, $\mathcal{I}^c$ denotes its complement, and $\left | \mathcal{I} \right | $ represents its cardinality. If $X$ follows a Bernoulli distribution with $P(X=1)=p$, we denote it by $X\sim Ber(p)$. The capacity of channel $W$ is denoted by $C(W)$. We denote the binary logarithm and natural logarithm by $\log$ and $\ln$, respectively, and information is measured in bits.

\section{Preliminaries}

\subsection{Adaptive Steganographic Model}
We use $X$ to denote the cover random variable and use $Y$ to denote the stego random variable after embedding. 
The stego sequence $y_1^N$ is obtained by embedding the secret message sequence $m_1^q$ into the cover sequence $x_1^N$. Typically, images are used as data carriers for both the cover and stego. In this paper, we consider binary embedding, hence $x_1^N, \!y_1^N$, and $m_1^q$ are binary sequences, with $x_i$ and $y_i$ being the Least Significant Bit (LSB) of the $i$-th pixel of the cover and stego, respectively. Note that the $q$-ary embedding can be implemented using multi-layered binary embedding \cite{filler2011minimizing}, \cite{filler2010using}.

We define the embedding modification probability for $x_i$ as $p_i \triangleq P(y_i \neq x_i \! \mid \! x_i)$. The distortion caused by embedding is measured by Hamming distortion, with a positive weight $\rho_i$. In additive distortion model, the total distortion is the sum of the distortion of each pixel given by
\begin{align}
D\left(x_1^N,y_1^N\right) & = \sum_{i  = 1}^{N}{\rho_i\cdot (x_i\oplus y_i)},
\end{align}
where $\oplus$ denotes the XOR operation. The formulations of the PLS and DLS problems are introduced in the following.

1) PLS: given a fixed number of embedded message bits $q$, minimize the total average distortion: 
\begin{align}
& \underset{p_1^N}{\text{minimize}} \quad E(D) = \sum_{i = 1}^{N} p_i \rho_i \label{eq:PLSmin}, 
\\
& \text{subject to} \quad H(p_1^N) = \sum_{i = 1}^{N} h_2(p_i) = q \label{eq:PLSconstraint}.
\end{align}

2) DLS: given a fixed total average distortion $D_\epsilon$, maximize the number of embedded message bits: 
\begin{align}
& \underset{p_1^N}{\text{maximize}} \quad H(p_1^N) = \sum_{i=1}^{N} h_2(p_i) \label{eq:DLSmin},  \\
& \text{subject to} \quad E(D) = \sum_{i=1}^{N} p_i \rho_i = D_\epsilon \label{eq:DLSconstraint}.
\end{align}

For the PLS problem, the optimal embedding has the form of a Gibbs distribution \cite{filler2010gibbs}:
\begin{align}
P_\lambda\left(y_i|x_i\right) & = \frac{\exp{\left(-\lambda\rho_i\cdot\left(x_i\oplus y_i\right)\right)}}{\sum_{t_i\in \left \{ 0, 1 \right \} }\exp{\left(-\lambda\rho_it_i\right)}},\ 1\le i\le N, \label{eq:gibbs-distribution}
\end{align}
where $\lambda \in [0, \infty)$ can be determined by \eqref{eq:PLSconstraint}. In practice, $H(p_1^N)$ is a monotonically decreasing function of $\lambda$, which can be found using a simple binary search. The PLS and DLS problems are dual problems of each other. The optimal embedding distribution for the DLS problem is also given by \eqref{eq:gibbs-distribution}. In this case, $\lambda$ can be determined in the same way by \eqref{eq:DLSconstraint}.

\subsection{Polar Codes}
For channel polarization, $N$ independent copies of a given B-DMC $W$ are combined into a vector channel $W_N$, which is split into $N$ subchannels  $W_N^{(i)}$, $1 \le i \le N$ \cite{arikan2009channel}. As $N$ increases to infinity, almost all subchannels have capacity close to 0 or 1, and the proportion of the subchannels with capacity close to 1 approaches $C(W)$.

\subsubsection{Polar Encoding}
Polar codes can be identified by a parameter tuple $(N,K,\mathcal{A},u_{\mathcal{A}^c})$, where $N=2^n$ for a positive integer. $\mathcal{A}$ is the set of indices for the $K$ information bits, and $u_{\mathcal{A}^c}$ represents the $N-K$ frozen bits.

The construction of $(N,K)$ polar codes refers to selecting the $K$ most reliable subchannels to transmit the information bits $u_{\mathcal{A}}$. One usually uses the Bhattacharyya parameter \cite{arikan2009channel} to evaluate the reliability of the subchannels. 
The Bhattacharyya parameter of $i$-th subchannel $W_N^{(i)}$ is defined as
\begin{align*}
Z(W_N^{(i)})\triangleq \!\sum_{\substack{(y_1^N, u_1^{i-1}) \in \! \\ \! \mathcal{Y}^N \!\times \!\{0,1\}^{i-1}}}\!\sqrt{W_N^{(i)}\!\left(y_1^N,\!u_1^{i-1}\middle|0\right)W_N^{(i)}\!\left(y_1^N,\!u_1^{i-1}\middle|1\right)}.
\end{align*}
For binary symmetric channel (BSC), $Z(W_N^{(i)})$ can be recursively approximated by Equivalent Bhattacharyya Parameter Construction \cite{arikan2008performance}. 
The degrading merging algorithm \cite{tal2013construct} is more commonly used, as it can estimate $Z(W_N^{(i)})$ with arbitrary precision.

After acquiring the information indices $\mathcal{A}$, the generator matrix is calculated by $G_N=B_N\left[\begin{smallmatrix}1&0\\1&1\end{smallmatrix}\right]^{\otimes n}$, where $B_N$ is known as the \emph{bit-reversal} permutation matrix and $\otimes$ denotes the Kronecker product. For encoding, $u_{\mathcal{A}^c}$ and $u_\mathcal{A}$ are combined into the vector $u_1^N = [u_{\mathcal{A}^c}, u_\mathcal{A}]$, where $u_\mathcal{A}$ represents the information bits and  $u_{\mathcal{A}^c}$ represents the frozen bits known to both the encoder and decoder. Then, the codeword is given by $x_1^N = u_1^N G_N$.

\subsubsection{Polar Decoding}
Consider the polar code represented by $(N,K,\mathcal{A},u_{\mathcal{A}^c})$. The decoder's task is to generate an estimate $\hat{u}_\mathcal{A}$ of $u_\mathcal{A}$ from the received $y_1^N$. 
The Successive Cancellation (SC) decoder, introduced by Ar{\i}kan \cite{arikan2009channel}, uses Log-Likelihood Ratio (LLR) to decide the value of $\hat{u}_i(i\in\mathcal{A})$ sequentially. The LLR can be calculated recursively from equations (75) and (76) presented in \cite{arikan2009channel}.

The Successive Cancellation List (SCL) \cite{balatsoukas2015llr,chen2012list} decoder is an enhancement based on the SC decoder. The SCL decoder extends the single decoding path of the SC decoder to $L$ paths. 
In the decoding process, the SCL decoder retains at most $L$ decoding paths with the best path metrics at each step. Upon completion of decoding, the SCL decoder selects the most reliable path as the output of the decoder. The complexities of the SC and SCL decoders are $O(N \log N)$ and $O(L \cdot N \log N)$, respectively. The notation $\text{SCL}(y_1^N, W^N, u_{\mathcal{A}^c}, \mathcal{A}^c, L)$ represents the decoding estimate of an SCL decoder with list size $L$, given the output $y_1^N$, channel $W^N$, frozen bits $u_{\mathcal{A}^c}$ and frozen indices $\mathcal{A}^c$.

\section{Polar Codes for adaptive steganography}
\begin{figure}[ht]
    \centering
    \includegraphics[width=9cm]{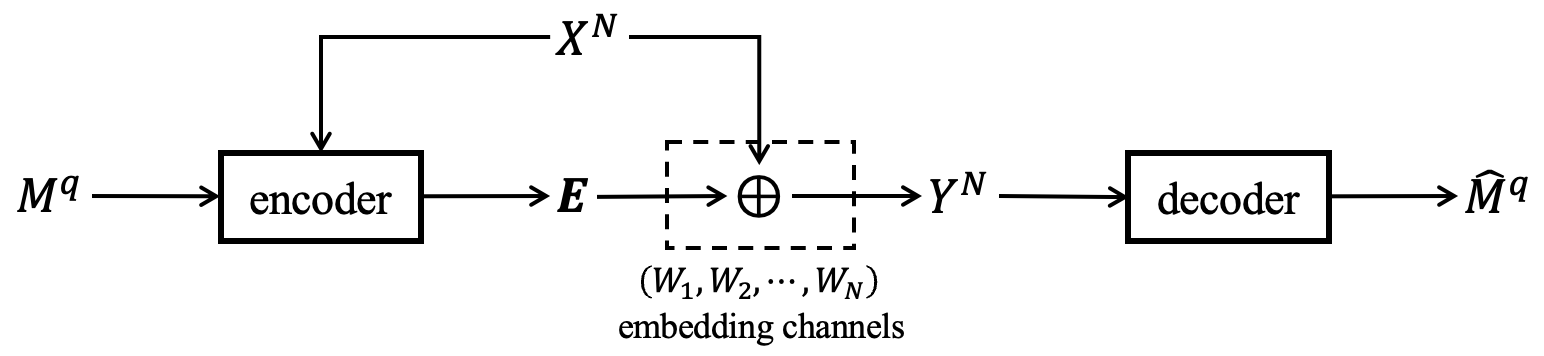}
    \caption{Illustration of the adaptive steganography.}
    \label{fig:non-robust}
\end{figure}

Let \( M \) represent the secret message. Embedding \( M^q \) into \( X^N \) can be viewed as passing \( X^N \) through \( (W_1, W_2, \cdots, W_N) \), where \( W_i \) is a BSC with the crossover probability of \( p_i \), denoted by \( \text{BSC}(p_i) \), representing the \( i \)-th embedding channel. The embedding capacity of \( W_i \) is complementary to the channel capacity, given by \( h_2(p_i) \). The adaptive steganography process is shown in Fig. \ref{fig:non-robust}. Let $\mathbf{E} = (E_1, E_2, \ldots, E_N)$, where \( E_i \) is the RV corresponding to \( W_i \) and \( E_i \sim Ber(p_i) \). Notice that the cover $X$ is a BSS and the stego $Y$ is also a BSS since the embedding channel is symmetric, demonstrating the distribution-preserving property of the steganography. Based on the aforementioned symmetry, $X_i$ can also be considered as obtained from $Y_i$ through $W_i$.
The channel polarization of \( (W_1, W_2, \cdots, W_N) \) is represented as \( (W_1, W_2, \cdots, W_N) \mapsto (W_N^{(1)}, W_N^{(2)}, \cdots, W_N^{(N)}) \).

For \( 0 < \beta < \frac{1}{2} \), partition \([N]\) into
\begin{align*}
\mathcal{F} \triangleq \left\{ i \in [N] : Z(W_N^{(i)}) \geq 1 - 2^{-N^\beta} \right\} 
\end{align*}
and \(\mathcal{I} = \mathcal{F}^c\), where \(\mathcal{F}\) is the indices corresponding to \( M^q \).
Equivalent Bhattacharyya Parameter Construction can be generalized to the case of parallel channels \cite{chen2013practical}. Therefore, \( Z(W_N^{(i)}) \) can still be recursively approximated.

For message embedding, we obtain \( u_\mathcal{F} \) by \( u_\mathcal{F} = m_1^q \). For each \( i \in \mathcal{I} \),
\begin{align}
\hat{u}_i & = \underset{u\in\{0, 1\}}{\rm argmax} \, P_{\hat{U}_i | \hat{U}_1^{i-1}, X_1^N} \left( u \mid \hat{u}_1^{i-1}, x_1^N \right).
\end{align}
Finally, we derive the stego sequence as \( y_1^N = u_1^N G_N \). Using the SCL decoder, we have
\begin{align}
\left(u_\mathcal{F},u_\mathcal{I}\right)  \!= u_1^N & \!= \text{SCL}\left(x_1^N\!,\left(W_1,W_2,\cdots\!,W_N\right)\!,m_1^q,\mathcal{F},L\right). \notag
\end{align}
The decoder calculates the LLR with the initial value \( L_1^{\left(1\right)}\left(x_i\right) = \left(1 - 2x_i\right) \ln{\frac{1 - p_i}{p_i}} \).

For message extraction, after receiving the stego sequence \( y_1^N \), the receiver uses the same construction method as the sender to obtain \( \mathcal{F} \) and \( \mathcal{I} \). Given \( G_N^{-1} = G_N \) for the 2-by-2 binary kernel \cite{arikan2009channel}, we have \( u_1^N = y_1^N G_N \), then \( m_1^q = u_\mathcal{F} \). 

Given a set of $N$ parallel channels $\{W_1, W_2, \cdots, W_N\}$, each component channel is used $J$ times. Let $w_{r,t}$ denote the $t$-th use of component channel $W_r$, that is, $w_{r,t}\Leftrightarrow W_r$, where the equivalence relation $\Leftrightarrow$ indicates the two channels have the same transition probability for $r=1,2,\cdots,N$ and $t=1,2,\cdots,J$. We use notation $w_1^{NJ}$ to denote a vector of $NJ$ independent channels, where $w_{i}$ is related to the channel uses $w_{r,t}$ with a one-to-one mapping
\begin{align}
    \pi:\{1,2,\cdots ,NJ\}\to \{1,2,\cdots N\} \times \{1,2,\cdots J\},
\end{align}
that is, $\pi(i)=(r,t)$, and
\begin{align}
    w_i\Leftrightarrow w_{\pi(i)} \Leftrightarrow w_{r,t} \Leftrightarrow W_r,
\end{align}
where $i \in \!\{1,\cdots\!,NJ\}$, $\!r \in \!\{1,\cdots\!,N\}$ and $\!t \in \!\{1,\cdots\!,J\}$.

By modifying the decision rule to the randomized rounding where $\hat{u}_i$ is obtained from $Ber(\frac{1}{1+\text{exp}(L_N^{(i)})})$, 
we can derive Theorem \ref{thm:1}, with its proof given in the Appendix. Likewise, Lemma \ref{lemma: independent uF} can be adapted from \cite{korada2009polar} Lemma 3.14, owing to the symmetry of the embedding channel and distortion function.

\begin{theorem}\label{thm:1}
Let the cover $X$ be a BSS. Consider any independent parallel embedding channel BSCs $\{W_1, W_2, \cdots, W_N\}$, where $N = 2^n$, for a fixed positive integer \(n\). For each embedding channel \(W_i\), the crossover probability is \(p_i \in [0, \frac{1}{2}]\). Using each of the $N$ channels $J$ times, where $J = 2^j$ and $j$ is a non-negative integer, we obtain $NJ$ channels $w_1^{NJ}$. 
Fix the design distortion $D= \frac{1}{N}\sum_{i = 1}^{N} p_i \rho_i$ and $0<\beta<\frac{1}{2}$. For any embedding rate  $R < \frac{1}{N}H(Y_1^N|X_1^N)= \frac{1}{N}\sum_{r = 1}^{N} h_2(p_r)$, there exists a sequence of polar codes of length $NJ$ with rates $R_{NJ}\ge R$ so that under SC encoding using randomized rounding they achieve expected distortion $D_{NJ}$ satisfying
\begin{align*}
    D_{NJ} \le D + O(2^{-(NJ)^\beta}).
\end{align*}
The encoding as well as decoding complexity of this scheme is $O(NJ\log{(NJ)})$.

\end{theorem}

\begin{lemma} \label{lemma: independent uF}
    The average distortion $D_{NJ}(F,u_F)$ is independent of the choice of $u_F\in\left\{0,1 \right\}^{|F|}$.
\end{lemma}

\section{Polar Codes for pixel-sensitive and \\ robust steganography}
In practical applications, the receiver may acquire noisy stego due to various reasons, such as noise attacks from an adversary or communication through a specific channel. For convenience, these are collectively called the \textit{attack channels} in this paper. The aforementioned adaptive steganography cannot resist noise, making it difficult to extract accurate secret messages. To protect secret messages from noise attacks, we propose pixel-sensitive and robust steganography. In this paper, we consider only noise that independently follows a Bernoulli distribution. Note that many other types of attack channels can be approximated as a BSC.

Equations \eqref{eq:PLSmin}-\eqref{eq:DLSconstraint} provide the forms of adaptive steganography. Here, we redefine the formulations in robust scenarios of adaptive steganography. The \(i\)-th attack channel \(Q_i\) is assumed to \(\text{BSC}(\theta_i)\), with the corresponding noise RV \(Z_i \sim Ber(\theta_i)\). Restate the PLS and DLS problem as follows.

1) PLS: 
\vspace{-0.4em}
\begin{align}
& \underset{p_1^N}{\text{minimize}} \quad E(D) = \sum_{i=1}^{N} p_i \rho_i \label{eq:PLS-robust-bound} , 
\\[-4pt]
& \text{subject to} \quad \sum_{i=1}^{N} h_2(p_i) - \sum_{i=1}^{N} h_2(\theta_i) = q. \label{eq:PLS-robust-le}
\end{align}

2) DLS:
\vspace{-0.3em}
\begin{align}
& \underset{p_1^N}{\text{maximize}} \quad H(p_1^N) = \sum_{i=1}^{N} h_2(p_i) - \sum_{i=1}^{N} h_2(\theta_i)  \label{eq:DLS-robust-embcapacity}, 
\\[-2pt]
& \text{subject to} \quad E(D) = \sum_{i=1}^{N} p_i \rho_i = D_\epsilon. 
\end{align}
It is evident that our scheme achieves noise robustness by sacrificing some embedding rate due to the additional attack.
\begin{figure}[ht]
    \centering
    \includegraphics[width=9cm]{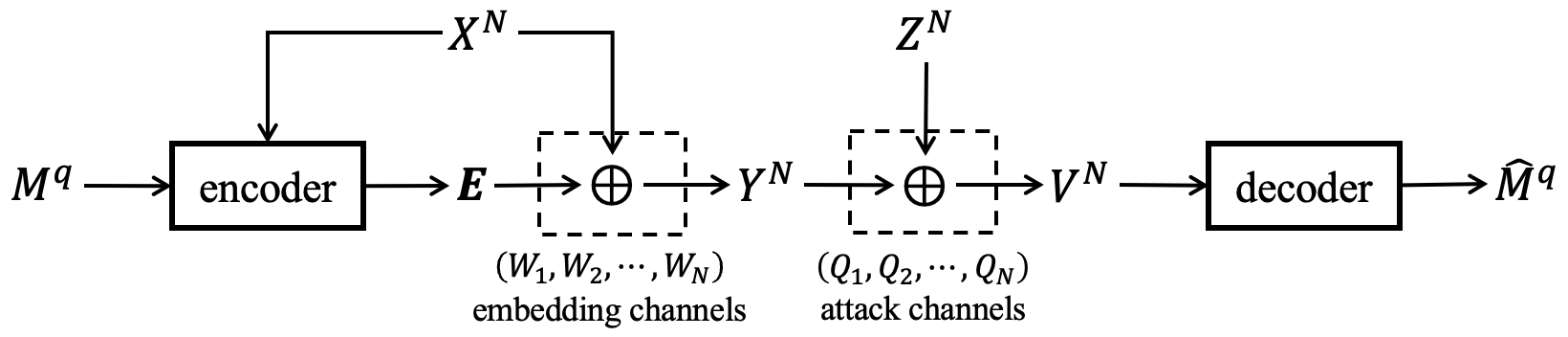}
    \caption{Illustration of the pixel-sensitive and robust steganography.}
    \label{fig:robust}
\end{figure}

Below, we present the pixel-sensitive and robust steganographic scheme based on polar codes. For \(0 < \beta < \frac{1}{2}\), the index sets are obtained as follows:
\begin{align}
\mathcal{F}_1&\triangleq\left\{i\in\left[N\right]:Z\left(W_N^{\left(i\right)}\right)\geq1-2^{-N^\beta}\right\}
\\
\mathcal{F}_2&\triangleq\left\{i\in\left[N\right]:Z\left(Q_N^{\left(i\right)}\right)\geq2^{-N^\beta}\right\}.
\end{align}
Next, we partition \(\left[N\right]\) into \(\mathcal{F} \triangleq \mathcal{F}_1 \cap \mathcal{F}_2\), \(\mathcal{I} \triangleq \mathcal{F}_1 \setminus \mathcal{F}_2\), and \(\mathcal{P} \triangleq \mathcal{F}_1^c\).
For the construction of parallel channels, the Monte Carlo construction proposed by Ar{\i}kan \cite{arikan2009channel}, which statistically estimates the decoding error rate of subchannels through experimental simulation, is more effective. The Monte Carlo construction algorithm is detailed in the pseudocode of Algorithm 1 in the Appendix. As the number of simulation runs increases, the statistical results become more effective in reflecting the reliability of the subchannels.

For message embedding, with the random frozen bits \(F \sim Ber\left(\frac{1}{2}\right)\), we obtain \(u_\mathcal{F}\) and \(u_\mathcal{I}\) by \(u_\mathcal{F} = f^{\left|\mathcal{F}\right|}\) and \(u_\mathcal{I} = m_1^q\), where $u_\mathcal{F}$ is a vector of frozen bits pre-shared between sender and receiver. For each \(i \in \mathcal{P}\), we compute
\begin{align}
\hat{u}_i & = \underset{u\in\{0,1\}}{\rm argmax} \, P_{\hat{U}_i | \hat{U}_1^{i-1}, X_1^N} \left( u \mid \hat{u}_1^{i-1}, x_1^N \right).
\end{align}
Finally, we derive the stego sequence as \(y_1^N = u_1^N G_N\). Using the SCL decoder, we have
\vspace{-0.4em}
\begin{multline}
\hspace{-1em}\left(u_\mathcal{F}, u_\mathcal{I}, u_\mathcal{P}\right) \!= u_1^N \!= \\
\hspace{0em}\text{SCL}\left(x_1^N\!, \left(W_1, W_2, \cdots\!, W_N\right)\!, (f^{\left|\mathcal{F}\right|}\!, m_1^q), \left(\mathcal{F}\!, \mathcal{I}\right)\!, L\right).
\notag
\end{multline}
The decoder calculates the LLR with the initial value 
$L_1^{\left(1\right)}\left(x_i\right)=\left(1-2x_i\right)\ln{\frac{1-p_i}{p_i}}$.

 For message extraction, after receiving the sequence \(v_1^N\) from attack channels \(\left(Q_1, Q_2, \ldots, Q_N\right)\), the receiver uses the same construction method as the sender to obtain \(\mathcal{F}_1\) and \(\mathcal{F}_2\), thus deriving \(\mathcal{F} \triangleq \mathcal{F}_1 \cap \mathcal{F}_2\), \(\mathcal{I} \triangleq \mathcal{F}_1 \setminus \mathcal{F}_2\), and \(\mathcal{P} \triangleq \mathcal{F}_1^c\). We obtain \(\hat{u}_\mathcal{F}=u_\mathcal{F}\). For each \(i \in \mathcal{I} \cup \mathcal{P}\),
 \begin{align}
\hat{u}_i & = \underset{u\in\{0,1\}}{\rm argmax} \, P_{\hat{U}_i | \hat{U}_1^{i-1}, V_1^N} \left( u \mid \hat{u}_1^{i-1}, v_1^N \right).
\end{align}
\(\hat{u}_\mathcal{I}\) is the extracted secret message. Using the SCL decoder, we have

\begin{align}
\left(u_\mathcal{F},{\hat{u}}_\mathcal{I},{\hat{u}}_\mathcal{P}\right)\! = {\hat{u}}_1^N\! & = \text{SCL}\left(v_1^N\!, \left(Q_1, Q_2, \cdots\!, Q_N\right)\!, u_\mathcal{F}\!, \mathcal{F}\!, L\right).
\notag
\end{align}
The decoder calculates the LLR with the initial value 
$L_1^{\left(1\right)}\left(v_i\right)=\left(1-2v_i\right)\ln{\frac{1-\theta_i}{\theta_i}}$.

Let  $W \preccurlyeq Q$  denote that channel $W$ is degraded with respect to channel $Q$, as defined in \cite{tal2013construct}. The following lemma can be easily adapted from \cite{tal2013construct}. For brevity, we omit the proof.

\begin{lemma}\label{lem:2}
For B-DMCs \(W_1, W_2, Q_1 \, \text{and} \, Q_2\), \(W^-\) and \(W^+\) are obtained from the polarization process \(\left(W_1, W_2\right) \mapsto \left(W^-, W^+\right)\), while \(Q^-\) and \(Q^+\) are obtained from the polarization \(\left(Q_1, Q_2\right) \mapsto \left(Q^-, Q^+\right)\). If \(W_1 \preccurlyeq Q_1\) and \(W_2 \preccurlyeq Q_2\), then
\begin{align}
W^-\preccurlyeq Q^-\ \ \text{and} \ \ W^+\preccurlyeq Q^+.
\end{align}
\end{lemma}
\begin{corollary}\label{cor:1}
For B-DMCs \(W_i \ \text{and} \  Q_i\) for \(i \in \left[N\right]\), $W_N^{\left(i\right)}$ and $Q_N^{\left(i\right)}$ are obtained from the polarization process $\left(W_1, W_2, \ldots, W_N\right) \mapsto (W_N^{\left(1\right)}, W_N^{\left(2\right)}, \ldots, W_N^{\left(N\right)})$ and $\left(Q_1, Q_2, \ldots, Q_N\right) \mapsto (Q_N^{\left(1\right)}, Q_N^{\left(2\right)}, \ldots, Q_N^{\left(N\right)})$, respectively. If \(W_i \preccurlyeq Q_i\) for \(i \in \left[N\right]\), then
\begin{align}
W_N^{\left(i\right)}\preccurlyeq Q_N^{\left(i\right)},\ \text{for} \ i\in\left[N\right].
\end{align}
\end{corollary}

\begin{theorem} \label{thm:2}
Let the cover $X$ be a BSS. Consider sets of independent parallel BSCs \(\{W_1, W_2, \cdots, W_N\}\) and \(\{Q_1, Q_2, \cdots, Q_N\}\) for embedding and attack channels, respectively, where \(N = 2^n\), for a fixed positive integer \(n\), and the crossover probability pair $(p_i, \theta_i)$ of each pair \((W_i, Q_i)\) satisfy \(0 \le \theta_i \le p_i \le \frac{1}{2}\). Using each pair of \(N\) channels \((W_i, Q_i)\) \(J\) times, where \(J = 2^j\) and \(j\) is a non-negative integer, we obtain \(NJ\) pairs of channels \((w_1^{NJ}, q_1^{NJ})\). 
Fix the design distortion $D= \frac{1}{N}\sum_{i = 1}^{N} p_i \rho_i$ and $0<\beta<\frac{1}{2}$. For any embedding rate  $R < \frac{1}{N}H(Y_1^N|X_1^N)-\frac{1}{N}H(Y_1^N|V_1^N)= \frac{1}{N}\sum_{r = 1}^{N} h_2(p_r)-\frac{1}{N}\sum_{r = 1}^{N} h_2(\theta_r)$, there exists a sequence of polar codes of length $NJ$ with rates $R_{NJ}\ge R$ so that under SC encoding with randomized rounding, they achieve expected distortion $D_{NJ}$ satisfying
\begin{align*}
    D_{NJ} \le D + O(2^{-(NJ)^\beta}).
\end{align*}
Futher, the block error probability satisfies
\begin{align*}
    P_{NJ}\le O(2^{-(NJ)^\beta}).
\end{align*}
The encoding as well as decoding complexity of this scheme is $O(NJ\log{(NJ)})$.
\end{theorem}

\begin{rem}
    Based on the assumptions of Theorem \ref{thm:2}, the proposed pixel-sensitive and robust steganography, in which the polar codes are constructed by the degrading merging algorithm\cite{tal2013construct}, can achieve the per-bit average embedding capacity in Equation \eqref{eq:DLS-robust-embcapacity} when $J$ and the limit of the output alphabet size are both sufficiently large. The expected distortion under SC encoding using randomized rounding and the block error probability under SC decoding are respectively upper bounded by \(D + O(2^{-(NJ)^\beta})\) and \(O(2^{-(NJ)^\beta})\), for \(0 < \beta < \frac{1}{2}\).
\end{rem}

\begin{rem}
    Similarly to Lemma \ref{lemma: independent uF}, the average distortion $D_{NJ}(F,u_F)$ and average block error probability $P_{NJ}(F,u_F)$ are independent of the choice of $u_F\in\left\{0,1 \right\}^{|F|}$, respectively.
\end{rem}

For cases not satisfying the conditions of Theorem \ref{thm:2}, \(\mathcal{F}_1\) and \(\mathcal{F}_2\) lack a clear nested relationship. This issue could potentially be addressed by universal polar codes \cite{hassani2014universal,csacsouglu2016universal}, which represents a promising direction for future research. However, in practice, \(\frac{\left|\mathcal{F}_2 \setminus \mathcal{F}_1\right|}{N}\) is typically small. Moreover, a larger \(\frac{\left|\mathcal{F}_2 \setminus \mathcal{F}_1\right|}{N}\) is associated with reduced robustness. 

We analyze two attack models below. The first model involves scenarios where the attacker launches attacks with the same intensity, or where the stego is transmitted over a channel. The second model describes scenarios where the attacker launches hidden linear attacks. The more complex the image texture surrounding a pixel, the better concealed higher-intensity attacks on that pixel become.

1) Attack Model 1 (AM1): Assume \(Z_i \sim Ber(\theta)\), $i\in[N]$, i.e., the attack model for each $i$ is fixed. The equation \eqref{eq:PLS-robust-le} becomes
$\sum_{i = 1}^{N} h_2(p_i) - Nh_2(\theta) = q.$
We ensure that \(\theta\) is known to both the sender and receiver. Thus, the optimal embedding distribution of $p_i$ remains as in \eqref{eq:gibbs-distribution}. Clearly, we cannot guarantee that \(p_i \geq \theta\) for all \(i \in \left[N\right]\).

2) Attack Model 2 (AM2): Assume \(R_a = \frac{\theta_i}{p_i} \in \left[0, 1\right]\) is a constant. We have \(p_i \geq \theta_i\) and \(Z_i \sim Ber(R_a p_i)\). The equation \eqref{eq:PLS-robust-le} becomes
$\sum_{i = 1}^{N} h_2(p_i) - \sum_{i = 1}^{N} h_2(R_a p_i) = q.$
If \(R_a\) is directly known to both the sender and receiver, solving for the optimal embedding distribution of $p_i$ becomes difficult. Therefore, to simplify the problem, we propose a two-phase approach, initially assuming that \(R_a\) is unknown, based on the reasonable assumption that the embedding loss due to robustness in AM2 is the same as in AM1. The proposed approach consists of the following phases:
\begin{itemize}
    \item \textbf{Phase 1:} Based on our assumption, we set 
    $\sum_{i=1}^{N} h_2(R_a p_i) = N h_2(\theta),$
    where \(\theta\) is known to both the sender and receiver, and \(R_a\) is a variable to be solved. Consequently, the equation \eqref{eq:PLS-robust-le} simplifies to 
    $\sum_{i = 1}^{N} h_2(p_i) - N h_2(\theta) = q.$
    At this point, solving for the optimal embedding distribution of \(p_i\) reverts to solving under AM1;

    \item \textbf{Phase 2:} With the optimal embedding distribution of \(p_i\) already known, we determine \(R_a\) by solving 
    $\sum_{i=1}^{N} h_2(R_a p_i) = N h_2(\theta).$
    Thus, \(\theta_i = R_a p_i\).
\end{itemize}

In fact, AM1 and AM2 yield the same optimal distribution of $p_i$ for the same problem, albeit with different noise parameters. Since AM2 satisfies the conditions in Theorem \ref{thm:2}, our scheme is validated. As for AM1, the effectiveness of our scheme will be assessed experimentally in the next section.

\section{SIMULATION EXPERIMENTS}
Next, we will conduct two simulation experiments. The first is to verify that the proposed pixel-sensitive and robust steganographic scheme can approach the theoretical bounds, and the second is to demonstrate that the scheme exhibits good robustness.
\subsection{Per-Bit Average Embedding Distortion}
We choose the PLS form for message embedding under AM1. The theoretical bound, \(E_b = \frac{1}{N} \sum_{i=1}^{N} p_i \rho_i\), is given by \eqref{eq:PLS-robust-bound}. We denote the distortion profile by the function $c(\cdot)$. Three common distortion profiles are considered: the constant profile \(c\left(x\right) = 1\), linear profile \(c\left(x\right) = x\), and square profile \(c\left(x\right) = x^2\). The value of $\rho_i$ is then computed as $c\left(\frac{i}{N}\right)$. The embedding rate \(R = \frac{q}{N}\) is measured in bit per pixel (bpp).

\begin{figure}[ht]
    \centering
    \includegraphics[width=7cm]{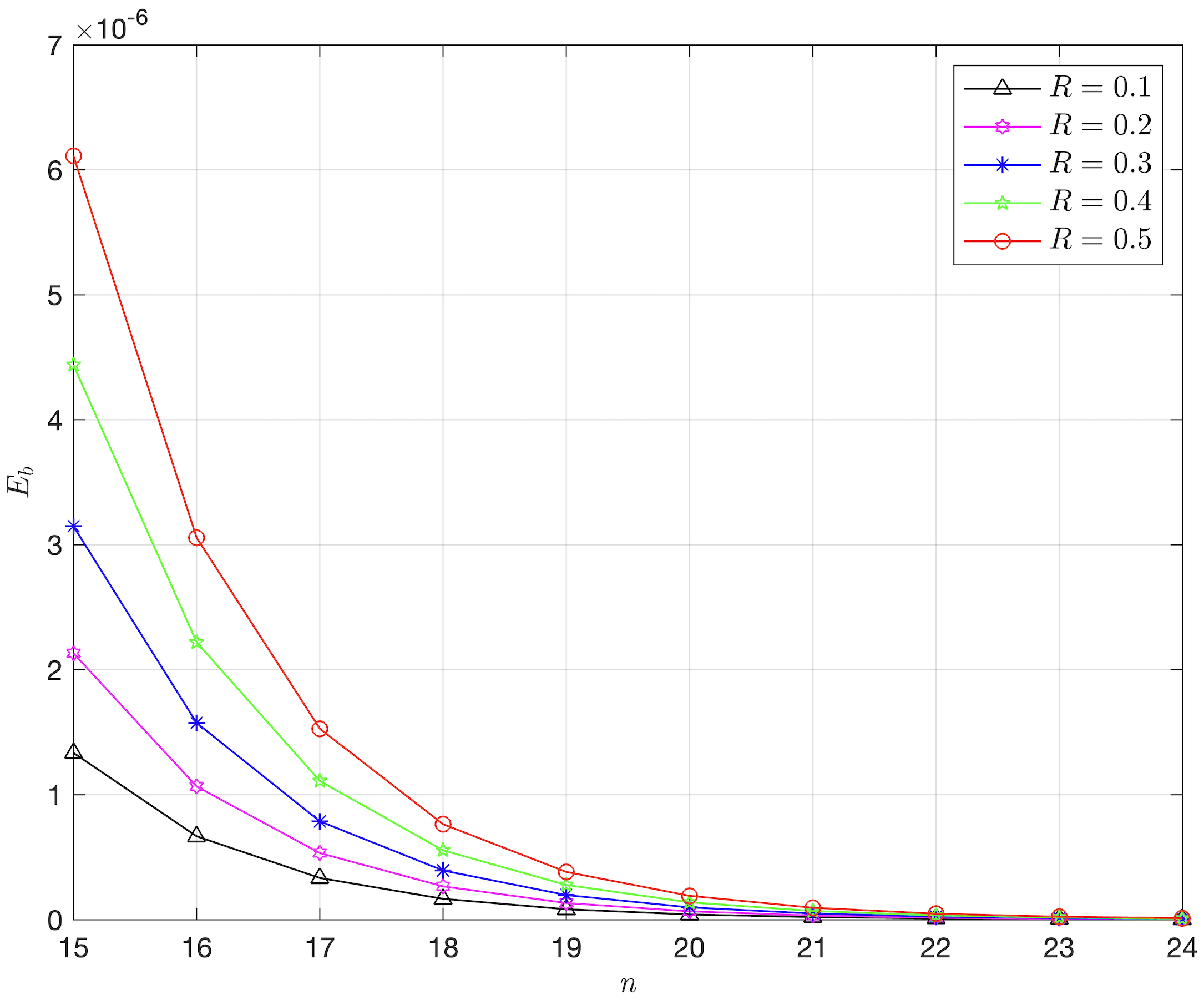}
    \caption{\(E_b\) variation with \(N = 2^n\) under linear distortion profile.}
    \label{fig:Eb}
\end{figure}

\begin{figure*}[t]
\centering
\subfigure[Linear distortion profile]
{
\begin{minipage}{0.43\textwidth}
\centering
\includegraphics[width=\textwidth]{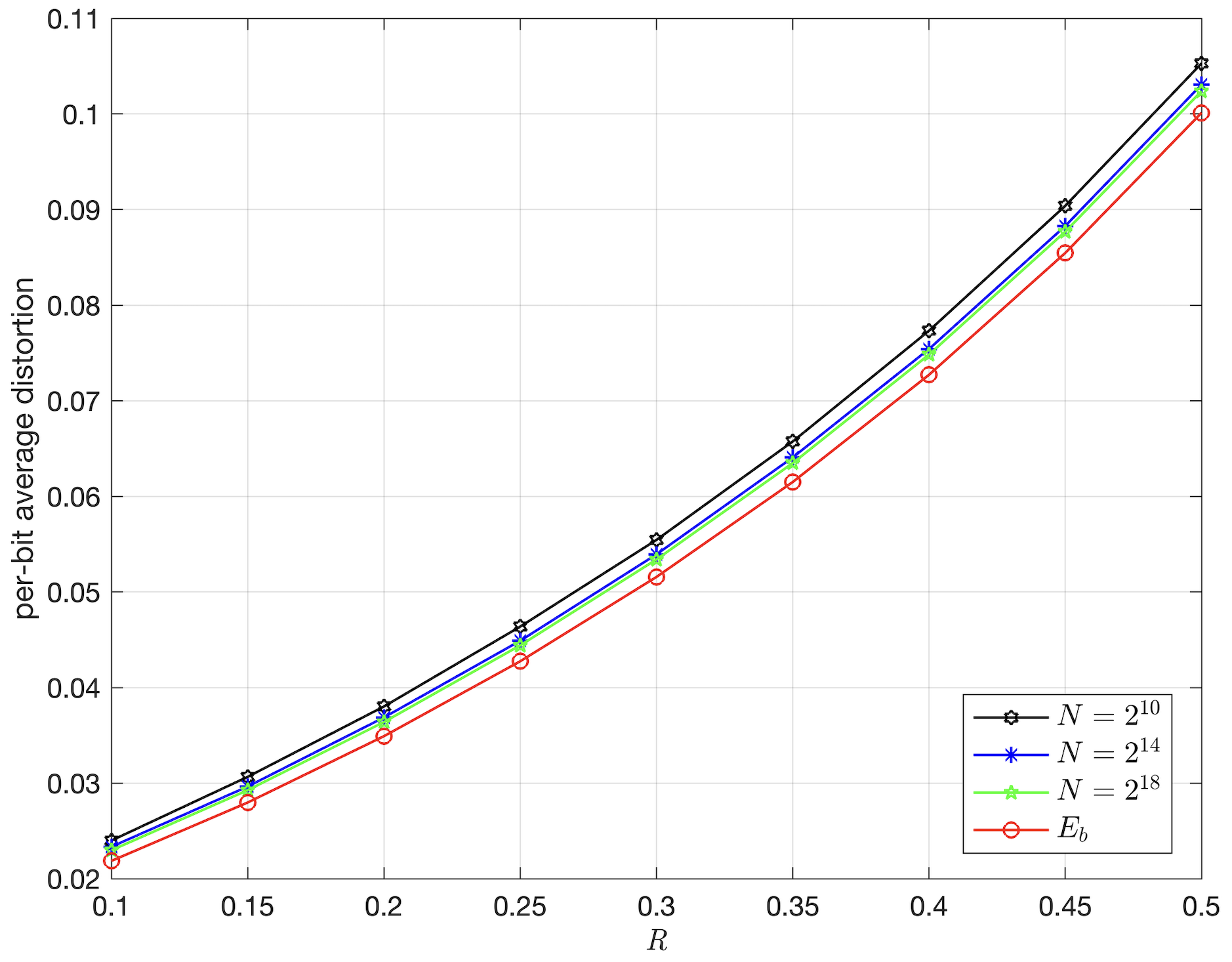}
\end{minipage}
}
\hfill
\subfigure[Square distortion profile]
{
\begin{minipage}{0.43\textwidth}
\centering
\includegraphics[width=\textwidth]{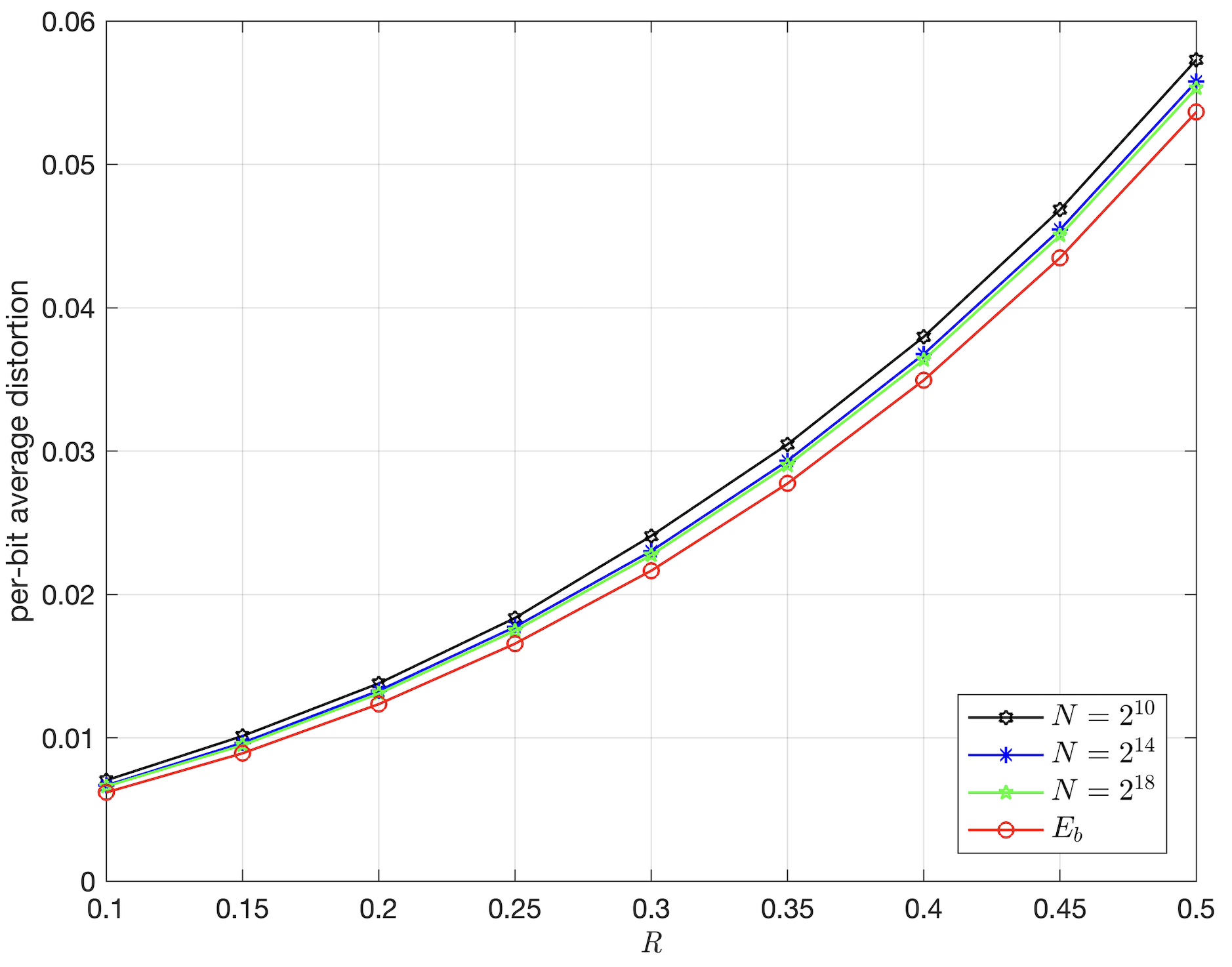}
\end{minipage}
}
\caption{Per-bit average distortion.}
\label{fig:testb}
\end{figure*}

The results for the constant distortion profile are provided in \cite{li2021robust}. Here, we consider the linear and square  distortion profiles. For both linear and square distortion profiles, \(E_b\) is monotonically non-increasing as \(N\) increases. This phenomenon is illustrated in Fig. \ref{fig:Eb}. As \(N\) increases, \(E_b\) converges to a fixed value. Therefore, we use \(E_b\) at \(N = 2^{22}\) as the theoretical bound in our experiments. With \(\theta = 0.05\) and \(R\) varying from 0.1 to 0.5, we use the SCL decoder with \(L = 16\), and the results are averaged over 100 simulations. The simulation results are shown in Fig. \ref{fig:testb}. It shows that the performance achieved by polar codes approaches the theoretical bound as \(N\) increases.

\subsection{Robustness}
Recall that \(\theta\) is the preset attack noise parameter for designing polar codes, not the actual attack noise parameter. For AM1, the actual attack noise \(\widetilde{Z}_i \sim Ber(\widetilde{\theta})\), and for AM2, \(\widetilde{Z}_i \sim Ber({\widetilde{\theta}}_i)\), $i\in[N]$. We define the constant \(R_{\theta=b} = \frac{{\widetilde{\theta}}_i}{\theta_i}\), where \(\theta_i\) is the value determined under AM2 for \(\theta=b\). The relationship between the actual noise $\widetilde{\theta}_i$ and the preset noise \(\theta_i\) can be characterized as follows:
\begin{itemize}
    \item \( R_{\theta=b} \!< \!1 \): the actual noise is less than the preset noise;
    \item \( R_{\theta=b} \!= \!1 \): the actual noise is equal to the preset noise;
    \item \( R_{\theta=b} \!> \!1 \): the actual noise is greater than the preset noise.
\end{itemize}
For \(N=2^{16}\) and \(R=0.1\), the simulations of our scheme using an SCL decoder with \(L=16\) are averaged over 200 runs. The simulation results are shown in Tables \ref{tab:AM1} and \ref{tab:AM2}, where '0' denotes an average bit error rate of less than \(10^{-5}\). Under the constant distortion profile, AM1 and AM2 are identical, thus omitted in Table \ref{tab:AM2}. The adaptive steganography performs poorly across the three common distortion profiles. In contrast, our proposed pixel-sensitive and robust steganographic scheme shows improved robustness against noise attacks, with robustness increasing as the preset attack noise \(\theta\) increases.

\begin{table}[h]
\centering
\caption{Average Secret Message Bit Error Rate (AM1)}
\begin{tabularx}{\columnwidth}{c|c|>{\centering\arraybackslash}X>{\centering\arraybackslash}X>{\centering\arraybackslash}X>{\centering\arraybackslash}X}
\toprule
Profiles & $\widetilde{\theta}$ & 0.001 & 0.002 & 0.004 & 0.006 \\
\midrule
\multirow{3}{*}{Constant} & \scriptsize{Adaptive} & \scriptsize{0.19023} & \scriptsize{0.48943} & \scriptsize{0.49883} & \scriptsize{0.49962} \\
 & \scriptsize{Ours($\theta=0.005$)} & \scriptsize{\textbf{0.00000}} & \scriptsize{0.00062} & \scriptsize{0.12212} & \scriptsize{0.49782} \\
 & \scriptsize{Ours($\theta=0.01$)} & \scriptsize{\textbf{0.00000}} & \scriptsize{\textbf{0.00000}} & \scriptsize{\textbf{0.00011}} & \scriptsize{\textbf{0.00123}} \\
\midrule
\multirow{3}{*}{Linear} & \scriptsize{Adaptive} & \scriptsize{0.16633} & \scriptsize{0.48138} & \scriptsize{0.49581} & \scriptsize{0.49888} \\
 & \scriptsize{Ours($\theta=0.005$)} & \scriptsize{\textbf{0.00000}} & \scriptsize{0.00017} & \scriptsize{0.11720} & \scriptsize{0.49678} \\
 & \scriptsize{Ours($\theta=0.01$)} & \scriptsize{\textbf{0.00000}} & \scriptsize{\textbf{0.00000}} & \scriptsize{\textbf{0.00003}} & \scriptsize{\textbf{0.00041}} \\
\midrule
\multirow{3}{*}{Square} & \scriptsize{Adaptive} & \scriptsize{0.15917} & \scriptsize{0.48110} & \scriptsize{0.49573} & \scriptsize{0.49901} \\
 & \scriptsize{Ours($\theta=0.005$)} & \scriptsize{\textbf{0.00000}} & \scriptsize{0.00036} & \scriptsize{0.13071} & \scriptsize{0.49645} \\
 & \scriptsize{Ours($\theta=0.01$)} & \scriptsize{\textbf{0.00000}} & \scriptsize{\textbf{0.00000}} & \scriptsize{\textbf{0.00003}} & \scriptsize{\textbf{0.00071}} \\
\bottomrule
\end{tabularx}
\label{tab:AM1}
\end{table}
\begin{table}[h]
\centering
\caption{Average Secret Message Bit Error Rate (AM2)}
\begin{tabularx}{\columnwidth}{c|c|>{\centering\arraybackslash}X>{\centering\arraybackslash}X>{\centering\arraybackslash}X>{\centering\arraybackslash}X}
\toprule
Profiles & $R_{\theta=0.005}$ & 0.2 & 0.4 & 0.8 & 1.2 \\
\midrule
\multirow{3}{*}{Linear} & \scriptsize{Adaptive} & \scriptsize{0.46863} & \scriptsize{0.49109} & \scriptsize{0.49805} & \scriptsize{0.49973} \\
 & \scriptsize{Ours($\theta=0.005$)} & \scriptsize{\textbf{0.00000}} & \scriptsize{0.00064} & \scriptsize{0.17693} & \scriptsize{0.49782} \\
 & \scriptsize{Ours($\theta=0.01$)} & \scriptsize{\textbf{0.00000}} & \scriptsize{\textbf{0.00000}} & \scriptsize{\textbf{0.00002}} & \scriptsize{\textbf{0.00039}} \\
\midrule
\multirow{3}{*}{Square} & \scriptsize{Adaptive} & \scriptsize{0.46857} & \scriptsize{0.49101} & \scriptsize{0.49836} & \scriptsize{0.49964} \\
 & \scriptsize{Ours($\theta=0.005$)} & \scriptsize{0.00008} & \scriptsize{0.00058} & \scriptsize{0.17746} & \scriptsize{0.49892} \\
 & \scriptsize{Ours($\theta=0.01$)} & \scriptsize{\textbf{0.00000}} & \scriptsize{\textbf{0.00000}} & \scriptsize{\textbf{0.00000}} & \scriptsize{\textbf{0.00044}} \\
\bottomrule
\end{tabularx}
\label{tab:AM2}
\end{table}

\section{Conclusion}
In this paper, we introduce adaptive steganographic schemes based on polar codes and present the embedding capacity-achieving theorem for the adaptive distortion. We redefine the mathematical formulations of the PLS and DLS problem in robust scenarios of adaptive steganography, propose a corresponding steganographic scheme based on polar codes, and present the embedding capacity-achieving theorem in these robust scenarios. Additionally, We describe two attack models and their solutions. We show our proposed scheme can approach the theoretical bounds and demonstrate strong robustness. Furthermore, this scheme is applicable to other attack and steganographic models as well.

\appendix
\subsection{Proof of Theorem 1}
Prior to the proof of the theorem, essential notations must be formally defined. In the channel polarization of \( (W, W, \cdots, W) \mapsto (W_N^{(1)}, W_N^{(2)}, \cdots, W_N^{(N)}) \), $b_1,\cdots,b_n$ denote the $n$-bit binary expansion of $i$ and $W_{(b_1,\cdots,b_n)} \triangleq W_N^{(i)}$. Let $\{B_n:n\ge1\}$ be a sequence of i.i.d. symmetric Bernoulli RVs defined over a probability space $(\Omega ,\mathcal{F},P)$. Let $\mathcal{F}_0=\{\phi, \Omega\}$ denote the trivial $\sigma$-field and let $\{\mathcal{F}_n,n\ge1\}$ denote the $\sigma$-fields generated by the RVs $(B_1,\cdots, B_n)$. Assume that $\mathcal{F}$ is such that $\mathcal{F}_0\subseteq \mathcal{F}_1\subseteq \cdots \subseteq \mathcal{F}$. Let $W_0=W$ and $\{W_n, n\ge0\}$ denote a tree process. We are required to use the Bhattacharyya parameters random process $\{Z_n:n\ge 0\}\colon=\{Z(W_n):n\ge0\}$.
\begin{proof}
Consider a specific channel mapping $\pi(i) = r+(t-1)N, \!r\in[N], \!t\in[J]$. The polarization of $NJ$ channels $w_1^{NJ}$ is divided into two stages. The first stage is the polarization of parallel channels. For each $t\in[J]$, the parallel polarization of $N$ channels is denoted by $(w_{1,t},w_{2,t},\cdots,w_{N,t}) \mapsto (W_N^{(1)}, W_N^{(2)}, \cdots, W_N^{(N)})$. The second stage is the polarization of identical channels. For each $r\in[N]$, the polarization of $J$ uses of  $W_N^{(r)}$ is denoted by $(W_N^{(r)},W_N^{(r)},\cdots,W_N^{(r)}) \mapsto (W_{NJ}^{(1+(r-1)J)}, W_{NJ}^{(2+(r-1)J)}, \cdots, W_{NJ}^{(rJ)})$. The $NJ$ subchannels $W_{NJ}^{(i)}(i\in[NJ])$ are obtained through two-stage polarization.

For the second-stage polarization, According to \cite{korada2009polar} Theorem 3.15, 
\begin{align*}
    \lim_{J=2^j, j\to \infty} Pr(Z_{j,r}\ge1-2^{-J^{\beta _r}}) = 1 - C(W_N^{(r)}),
\end{align*}
where $0<\beta_r<\frac{1}{2}, r\in[N]$ and $Z_{j,r}$ denotes the RV representing the identical channel polarization of $W_N^{(r)}$. Let $\beta'=min(\beta_1,\cdots, \beta_N)$, then
\begin{align*}
    \lim_{J=2^j, j\to \infty} Pr\left(Z_{nj}\ge1-2^{-J^{\beta'}}\right) &=  1 - \frac{1}{N}\sum_{r=1}^N C(W_N^{(r)}) \\ &=\frac{1}{N}\sum_{r = 1}^{N} h_2(p_r),
\end{align*}
where $Z_{nj}$ denotes the RV representing the channel polarization of $w_1^{NJ}$. Let $\beta=\frac{\ln{J}}{\ln{(NJ)}} \beta', 0<\beta\le\beta'<\frac{1}{2} $, then
\begin{align}
    \lim_{J=2^j, j\to \infty} Pr\left(Z_{nj}\ge1-2^{-(NJ)^{\beta}}\right) = \frac{1}{N}\sum_{r = 1}^{N} h_2(p_r). \label{eq:appendix, th1,polarization-eq}
\end{align}
According to parallel channel polarization theorem \cite{chen2013practical} Theorem 2, it is readily apparent that \cite{korada2009polar} Lemma 3.5-3.8 remain valid. 

    For $0<\beta<\frac{1}{2}$ and $\delta_{NJ}=\frac{1}{2NJd_{max}   }2^{-(NJ)^\beta}$, $d_{max}=\max_{r\in[N],\rho_r  \ne \infty  }\rho_r$, Consider a polar code with frozen set $F_{NJ}$,
\begin{align*}
    F_{NJ}=\left \{ i\in[NJ]:Z(W_{NJ}^{(i)})\ge 1-2\delta_{NJ}^2  \right \}.
\end{align*}
For $J$ sufficiently large there exists a $\beta'<\frac{1}{2}$ such that $2\delta_{NJ}^2>2^{-(NJ)^{\beta'}}$. Equation (\ref{eq:appendix, th1,polarization-eq}) implies that
\begin{align}
    \lim_{NJ=2^{n+j}, j\to \infty} \frac{\left | F_{NJ} \right | }{NJ} =\frac{1}{N}\sum_{r = 1}^{N} h_2(p_r).
\end{align}
The above equation implies that for any $\epsilon>0$ and for $J$ sufficiently large there exists a set $F_{NJ}$ such that
\begin{align*}
    \frac{\left | F_{NJ} \right | }{NJ} \ge \frac{1}{N}\sum_{r = 1}^{N} h_2(p_r)-\epsilon.
\end{align*}
From \cite{korada2009polar} Lemma 3.6, we know that
\begin{align}
    D_{NJ}(F_{NJ})\le D+2\left | F_{NJ} \right | d_{max}\delta_{NJ} \le D + O( 2^{-(NJ)^\beta} ), \label{eq: appendix, eq2}
\end{align} 
for $0<\beta<\frac{1}{2}$.
\end{proof}

Recall that $D_{NJ}(F_{NJ})$ is the average of the distortion over all choices of $u_{NJ}$. Since the average distortion fulfills (\ref{eq: appendix, eq2}) it follows that there must be at least one choice of $u_{NJ}$ for which
\begin{align*}
    D_{NJ}(F_{NJ},u_{F_{N\!J}})\le D + O ( 2^{-(NJ)^\beta} ), 0<\beta<\frac{1}{2}.
\end{align*}

\subsection{Proof of Theorem 2}
\begin{proof} 
Consider a specific channel mapping $\pi(i) = r+(t-1)N, \!r\in[N], \!t\in[J]$. We apply two-stage polarization to the $NJ$ channel \( q_1^{NJ} \). For the second-stage polarization, According to \cite{arikan2009rate} Theorem 1,
\begin{align*}
    \lim_{J=2^j, j\to \infty} Pr(Z_{j,r}<2^{-J^{\beta _r}}) = C(Q_N^{(r)}),
\end{align*}
where $0<\beta_r<\frac{1}{2}, r\in[N]$ and $Z_{j,r}$ denotes the RV representing the identical channel polarization of $Q_N^{(r)}$. Let $\beta'=min(\beta_1,\cdots, \beta_N)$, then
\begin{align*}
    \lim_{J=2^j, j\to \infty} Pr\left(Z_{nj}<2^{-J^{\beta'}}\right) &=  \frac{1}{N}\sum_{r=1}^N C(Q_N^{(r)}) \\ &=1-\frac{1}{N}\sum_{r = 1}^{N} h_2(\theta_r).
\end{align*}
where $Z_{nj}$ denotes the RV representing the channel polarization of $q_1^{NJ}$. Let $\beta=\frac{\ln{J}}{\ln{(NJ)}} \beta', 0<\beta\le\beta'<\frac{1}{2} $, then
\begin{align}
    \lim_{J=2^j, j\to \infty} Pr\left(Z_{nj}<2^{-(NJ)^{\beta}}\right) = 1-\frac{1}{N}\sum_{r = 1}^{N} h_2(\theta_r)
    \notag
\end{align}
and
\begin{align}
    \lim_{J=2^j, j\to \infty} Pr\left(Z_{nj}\ge2^{-(NJ)^{\beta}}\right) = \frac{1}{N}\sum_{r = 1}^{N} h_2(\theta_r).
    \label{eq:appendix, th2, channel polarization-eq}
\end{align}

    Let $\epsilon>0$ and $0<\beta<\frac{1}{2}$ be some constants. Let $\delta_{NJ}=\frac{1}{NJ}2^{-(NJ)^\beta}$. Let $\mathcal{F}_1$ and $\mathcal{F}_2$ denote the sets
    \begin{align}
    \mathcal{F}_1 &\triangleq\left\{i\in\left[NJ\right]:Z\left(W_{NJ}^{\left(i\right)}\right)\geq1-2^{-(NJ)^\beta}\right\}
    \notag
    \\
    \mathcal{F}_2&\triangleq\left\{i\in\left[NJ\right]:Z\left(Q_{NJ}^{\left(i\right)}\right)\geq2^{-(NJ)^\beta}\right\}.
    \notag
    \end{align}
    Equation (\ref{eq:appendix, th1,polarization-eq}) implies that for $J$ sufficiently large
    \begin{align*}
        \frac{\left | \mathcal{F}_1 \right | }{NJ} \ge \frac{1}{N}\sum_{r = 1}^{N} h_2(p_r)-\frac{\epsilon}{2}.
    \end{align*}
    Similarly, equation (\ref{eq:appendix, th2, channel polarization-eq}) implies that for $J$ sufficiently large
    \begin{align*}
        \frac{\left | \mathcal{F}_2 \right | }{NJ} \le \frac{1}{N}\sum_{r = 1}^{N} h_2(\theta_r)+ \frac{\epsilon}{2}.
    \end{align*}
    Since each pair \((W_r, Q_r)\) satisfy \(0 \le \theta_r \le p_r \le \frac{1}{2}\) for $r\in[N]$, we know
    \begin{align}
    W_{NJ}^{\left(i\right)}\preccurlyeq Q_{NJ}^{\left(i\right)},\ \text{for} \ i\in\left[NJ\right].
    \notag
    \end{align}
    When $J$ is sufficiently large,
    \begin{align*}
        \mathcal{F}_2 \subseteq \mathcal{F}_1.
    \end{align*}
     Partition \(\left[NJ\right]\) into \(\mathcal{F} \triangleq \mathcal{F}_1 \cap \mathcal{F}_2\), \(\mathcal{I} \triangleq \mathcal{F}_1 \setminus \mathcal{F}_2\), and \(\mathcal{P} \triangleq \mathcal{F}_1^c\).
    Therefore,
    \begin{align}
        \frac{\left | \mathcal{F} \right | }{NJ} &\le \frac{1}{N}\sum_{r = 1}^{N} h_2(\theta_r)+ \frac{\epsilon}{2},
    \\
        \frac{\left | \mathcal{I} \right |}{NJ} &\ge \frac{1}{N}\sum_{r = 1}^{N} h_2(p_r)-\frac{1}{N}\sum_{r = 1}^{N} h_2(\theta_r) - \epsilon, \label{embedding rate of robust}
    \\
        \frac{\left | \mathcal{P} \right | }{NJ} &< 1-\frac{1}{N}\sum_{r = 1}^{N} h_2(p_r)+\frac{\epsilon}{2}.
    \end{align}
    The embedding rate is $\frac{\left | \mathcal{I} \right |}{NJ}$, satisfing (\ref{embedding rate of robust}).

    For message embedding, the polar code constructed based on \(\mathcal{F}_1\) and \(\mathcal{I}_1 = \mathcal{F}_1^c\) is the same as that of the adaptive steganographic scheme. In conjunction with equation (\ref{eq: appendix, eq2}) and Lemma \ref{lemma: independent uF}, we know
    \begin{align}
    D_{NJ}(F_{NJ})\le D+2\left | F_{1} \right | \delta_{NJ} \le D + O ( 2^{-(NJ)^\beta} ).
    \end{align} 

    For message extraction, the polar code constructed based on \(\mathcal{F}_2\) and \(\mathcal{I}_2 = \mathcal{F}_2^c\) is a channel code for attack channels $q_1^{NJ}$. The block error probability satisfies
    \begin{align}
        P_{NJ} \le \sum_{i\in \mathcal{I}_2} Z(Q_{NJ}^{(i)}) \le O(2^{-(NJ)^\beta}).
    \end{align}
    
\end{proof}

\begin{algorithm}[ht!]
\caption{Monte Carlo Construction}
\label{alg: monte carlo constuction}
\begin{algorithmic}[1]
\Statex \textbf{Input:} code length $N$, information length $K$, BSCs \( (W_1, W_2, \cdots, W_N) \), simulation runs $T$
\Statex \textbf{Output:} information indices set $\mathcal{A}$, frozen indices set $\mathcal{A}^c$    
    
    \State define $e_1^N = 0_1^N$
    \For{$i = 1$ to $T$}
        \State $u_1^N \gets \text{randi}([0,1], 1, N)$
        \State $x_1^N = u_1^N G_N$
        \State generate noise $z_1^N$ from BSCs \( (W_1, W_2, \cdots, W_N) \)
        \State $y_1^N = x_1^N \oplus z_1^N$
        \State \parbox[t]{\dimexpr\linewidth-\algorithmicindent}{%
            calculate $L_{N}^{(i)}\left(y_{1}^{N}, \widehat{u}_{1}^{i-1}\right)$ using equations (75) and (76) from \cite{arikan2009channel}%
        }
        \If{$L_{N}^{(i)}\left(y_{1}^{N}, \widehat{u}_{1}^{i-1}\right) \ge 0$}
            \State $\hat{u}_i = 0$
        \Else
            \State $\hat{u}_i = 1$
        \EndIf
        \If{$u_i \neq \hat{u}_i$}
            \State $e_i = e_i + 1$
            \State $\hat{u}_i = u_i$
        \EndIf
    \EndFor
    \State Sort \( e_1^N \) in ascending order and store the indices in \( idx_1^N \)
    \State $\mathcal{A}=idx_1^K$, $\mathcal{A}^c=idx_{K+1}^N$
    \State \Return $(\mathcal{A}, \mathcal{A}^c)$
\end{algorithmic}
\end{algorithm}

\bibliographystyle{IEEEtran}
\bibliography{main}

\end{document}